\documentclass[preprint2]{emulateapj}

\slugcomment{version 2009 Jan 26: fm/deh}

\shorttitle{The Jet of 3C~17 and the Use of Jet Curvature as a Diagnostic
  of the X-ray Emission Process}
\shortauthors{Massaro, F., Harris, D. E., Chiaberge, M., Grandi, P., Macchetto, F. D., et al.}

\newcommand{\chn}{{\it Chandra}}

\begin{document}


\title{The Jet of 3C~17 and the Use of Jet Curvature as a Diagnostic
  of the X-ray Emission Process}


\author{F. Massaro\altaffilmark{1}, D. E. Harris\altaffilmark{1},
M. Chiaberge\altaffilmark{2,7}, P.~Grandi\altaffilmark{3},
F. D. Macchetto\altaffilmark{2}, S.~A.~Baum\altaffilmark{4},
C.~P.~O'Dea\altaffilmark{5} and A.~Capetti\altaffilmark{6}}



\altaffiltext{1}{Harvard, Smithsonian Astrophysical Observatory, 60 Garden Street, Cambridge, MA 02138}
\altaffiltext{2}{Space Telescope Science Institute, 3700 San Martine Drive, Baltimore, MD 21218}
\altaffiltext{4}{Carlson Center for Imaging Science 76-3144, 84 Lomb Memorial Dr., Rochester, NY 14623}
\altaffiltext{3}{INAF-IASF - Istituto di Astrofisica Spaziale e fisica cosmica di Bologna, Via P. Gobetti 101, 40129, Bologna, Italy}
\altaffiltext{5}{Dept of Physics, Rochester Institute of Technology, Carlson Center for Imaging Science 76-3144, 84 Lomb Memorial Dr., Rochester, NY 14623}
\altaffiltext{6}{INAF - Osservatorio Astronomico di Torino, Strada Osservatorio 20, I-10025 Pino Torinese, Italy}
\altaffiltext{7}{INAF - Istituto di Radioastronomia di Bologna, via Gobetti 101 40129 Bologna, Italy}

\begin{abstract} 
We report on the X-ray emission from the radio jet of 3C~17 from
\chn~observations and compare the X-ray emission with radio maps from
the VLA archive and with the optical-IR archival images from
the Hubble Space Telescope.  X-ray detections of two knots in the
3C~17 jet are found and both of these features have optical
counterparts.  We derive the spectral energy distribution for the
knots in the jet and give source parameters required for the various
X-ray emission models, finding that both IC/CMB and synchrotron are viable to explain the high energy emission.  
A curious optical feature (with
no radio or X-ray counterparts) possibly associated with the 3C~17 jet
is described.  We also discuss the use of curved jets for the problem of
identifying inverse Compton X-ray emission via scattering on CMB photons. 
\end{abstract}




\section{Introduction}
The X-ray radiation observed from radio jets is generally interpreted
to be from non-thermal processes, even if its nature is still unclear
for any particular jet.  It could be described in terms of synchrotron
emission or in terms of several varieties of inverse Compton
radiation.  So to understand the emission mechanisms related to these
components the multiwavelength approach is required.  If the X-ray
emission is synchrotron, electrons with Lorentz factors
$\gamma$ up to 10$^7$ are required whereas if the process is inverse
Compton radiation with seed photons due to the CMB (Tavecchio et al.,
2000), the X-rays would come from electrons with $\gamma\approx~100$
(Harris \& Krawczynski 2002).  To investigate the nature of the
emission in jets we analyze the jet of the powerful radio galaxy
3C~17.
 
3C~17 was observed during the first year of the \chn~3C snapshot
program, which started in AO-9 with 8ks observations of 30 of the
previously unobserved (by \chn) 3C sources with z$<$0.3.  
The 3C sample allows us to have multifrequency data
available from the HST and the VLA archives.  3C~17 is a radio galaxy
($z\sim$0.22, Schmidt et al. 1965) with a peculiar radio structure
investigated by Morganti et al. (1999).  Its H$\alpha$ emission has a
strong broad component and both the [O II] $\lambda$3727 and [O III]
$\lambda$5007 emission lines are extended (Dickson 1997).  This source
shows also a significant optical polarization in its nucleus
(Tadhunter et al. 1998), and its first detection in X-rays has been
reported by Siebert et al. (1996) using the ROSAT All Sky survey data.
 
Here, we report the major results concerning the multiwavelength
studies of the jet in 3C~17. We present the X-ray data of
this source together with the optical-IR images (HST) and the
radio maps (VLA archive).

For our numerical results, we use cgs units unless stated otherwise
and we assume a flat cosmology with $H_0=72$ km s$^{-1}$ Mpc$^{-1}$,
$\Omega_{M}=0.27$ and $\Omega_{\Lambda}=0.73$ (Spergel et al., 2007),
so 1$^{\prime\prime}$ is equivalent to 3.47 kpc.  Spectral indices,
$\alpha$, are defined by flux density, S$_{\nu}\propto\nu^{-\alpha}$.

\section{Observations and Data reduction}
\subsection{X-rays data}
3C~17 has been observed by Chandra (Obs ID 9292) on February 2, 2008, with the ACIS-S camera, 
operating in VFAINT mode, with an exposure of about 8 ksecs.
The data reduction has been performed following the standard
procedures described in the Chandra Interactive Analysis of
Observations (CIAO) threads
and using the CIAO software package v3.4.  The Chandra Calibration
Database (CALDB) version 3.4.2 was used to process all files.  Level 2
event files were generated using the $acis\_process\_events$ task,
after removing the hot pixels with $acis\_run\_hotpix$.  Events were
filtered for grades 0,2,3,4,6 and we removed pixel randomization.
Astrometric registration was done changing the appropiate keywords in
the fits header so as to align the nuclear X-ray position with that of
the radio.  We also registered the HST images in the same way.

We created 3 different fluxmaps (soft, medium, hard, in the ranges 0.5
-- 1, 1 -- 2, 2 -- 7 keV, respectively) by dividing the data with the
exposure maps. When constructing the fluxmaps, we normalized each
count by multiplying by h$\nu$ where $\nu$ corresponds to the energy
used for the corresponding exposure map.  Thus we can measure the flux
in any aperture in cgs units with only a small correction for the
ratio of the mean energy of the counts within the aperture to the
nominal energy for the band.

\begin{figure}[!htp]
\includegraphics[height=6cm,width=7cm,angle=0]{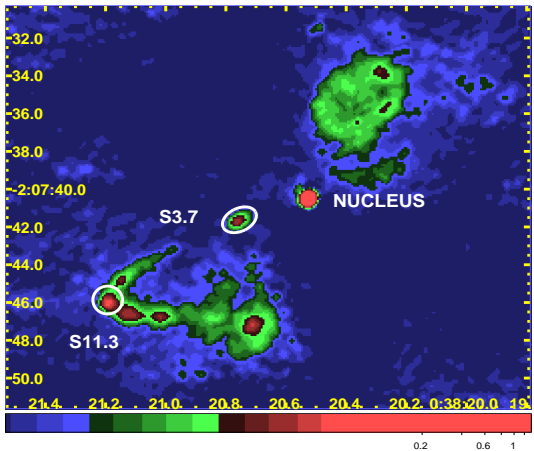}
\caption{The 5 GHz VLA map of Morganti et al. (1999) with a restoring
  beam of 0.4$^{\prime\prime}$.  The two knots of interest are
  emphasized (in white) with the regions used for
  photometry.}\label{fig:raff}
\end{figure}

Photometric apertures were constructed so as to accomodate the Chandra
point spread function and so as to include the total extent of the
radio structure.  They are shown in fig.~\ref{fig:raff}.  The
background regions have been chosen close to the source with
comparable size, typically two times bigger than the source region,
centered on a position where other sources or extended structure are
not present.  All X-ray flux densities have been corrected for the
Galactic absorption with the N$_H$ column densities given by Kalberla
et al. (2005), 2.86$\cdot$10$^{20}$cm$^{-2}$.

\subsection{Radio maps and HST images}
We compare our X-rays maps of 3C~17 with the VLA radio data described
in Morganti et al. (1999) at 4.8 GHz with a beamsize of
0.4$^{\prime\prime}$.  We also reduced archival VLA data at 1.54 GHz
and 14.9 GHz with AIPS standard reduction procedures.  The angular
resolution of these radio maps is $\approx$ 1.4$^{\prime\prime}$ and
the final image is in good agreement with the 4.8 GHz radio map.  The
amplitude calibrator used at 1.4 GHz was 3C 48 and the phase calibrator
was 0056-001. At 15GHz we used 0106+103 for both amplitude and
phase.

Concerning the other bands, we compared our X-ray image to the IR HST
observation at 1.6 $\mu$m (1.87$\cdot$10$^{14}$~Hz, H
band)\footnote{available on the {\it HST Snapshots of 3CR Radio
Galaxies} webpage, http://archive.stsci.edu/prepds/3cr/}, and to the
STIS visible image at 4.16$\cdot$10$^{14}$~Hz (7216 \AA, R band).  We
also used a FUV HST image at 1457\AA.

\section{Results}\label{sec:results}  
3C~17 is a broad lined radio galaxy (BLRG) (Buttiglione et
al. 2008) with a monochromatic radio luminosity, log P$_{1.4 GHz}
\approx$ 26.9 which is at least two orders of magnitude above the
division between FRI and FRII types (Ledlow \& Owen, 1996) but with an ambiguous 
radio morphology.  Miller \& Brandt (2008) provide a more
extensive discussion on classifying sources of this 'hybrid' type,
including 4C65.15, which is very similar to 3C~17 in many respects.

The nucleus of the host galaxy has been observed with the Very Long
Baseline Array by Venturi et al. (2000) who describe 3C~17 as a
``transition object'' between FRI and FRII.  The pc scale jet shows a
'core' with an extension in PA$\approx~100^{\circ}$ to 110$^{\circ}$,
followed by lower brightness features.  This position angle is
essentially the same as that of the first kpc scale knot, S3.7,
discussed below.

The kpc scale radio morphology is dominated by a single sided,
strongly curved jet (fig.~\ref{fig:raff}, as described by Morganti et
al. 1993, 1999), although there is lower brightness emission outside
the area covered in the figure.  Similarly to M87, 3C~17 was
originally classified as a 'core-halo' source.  The jet has a bright
knot at 3.7$^{\prime\prime}$ from the nucleus while the curved part
lies at about 11$^{\prime\prime}$ from the nucleus.

\subsection{Jet knots}

\begin{figure}[!htp]
\includegraphics[height=6cm,width=7cm,angle=0]{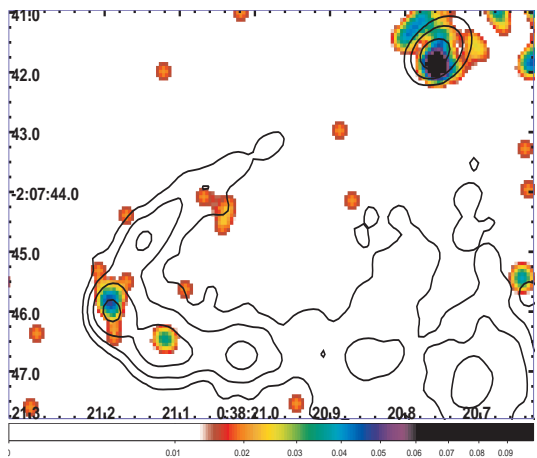}
\caption{The Chandra X-ray map.  Pixel randomization has been removed
  and the events between 0.5 and 7 keV were smoothed with a Gaussian
  of FWHM=0.87$^{\prime\prime}$.  The radio contours come from
  fig.~\ref{fig:raff} and start at 0.5 mJy/beam, increasing by factors
  of 4.  S3.7 is at the upper right and S11.3 to the lower left.}\label{fig:evt2}
\end{figure}

\begin{table}
\caption{Observed X-ray counts and fluxes for jet features.\label{tab:xflux}}
\begin{flushleft}
\begin{tabular}{lllll}
\hline
   & Soft & Medium  &  Hard  & Total \\
\hline 
\noalign{\smallskip}
Nominal Energy (keV) & 0.70 &1.4 & 4.0 \\
Band (keV) & 0.5-1.0 & 1-2 & 2-7 & 0.2-7 \\
\hline
\noalign{\smallskip}
S3.7  counts &  5 & 6 & 1 & 12 \\
S11.3 counts & 1 & 0 & 4 & 5\\
S3.7 flux  &  2.6$\pm$1.1 & 2.2$\pm$0.9 & 1.6$\pm$1.6 & 6.4 \\
S11.3 flux & 0.6$\pm$0.5 & 0 & 6.5$\pm$3.3 & 7.1\\
\noalign{\smallskip}
\hline
\end{tabular}\\
Flux units: 10$^{-15}$ erg~cm$^{-2}$~s$^{-1}$.

Notes to table:

{\footnotesize The average backgrounds measured  for the total 0.5-7 keV band from annular rings around the radio galaxy are 1.57 counts
(S3.7) and 0.08 counts (S11.3).}

\end{flushleft}
\end{table}

\begin{table}
\caption{Flux densities (cgs units) for 3C~17 knots\label{tab:fluxden}}
\begin{flushleft}
\begin{tabular}{lcrr}
\hline
Freq. (Hz)  & Band & S3.7    &   S11.3  \\
\hline 
\noalign{\smallskip}
1.66$\cdot$10$^{9}$      & L      & 80$\pm$10$\cdot10^{-26}$   & (190$\pm$10$\cdot10^{-26}$)  \\
4.86$\cdot$10$^{9}$      & C      &  30$\pm$1$\cdot10^{-26}$   & 83$\pm2\cdot10^{-26}$ \\
1.49$\cdot$10$^{10}$    & U       & 12.4$\pm$1$\cdot10^{-26}$  & (33$\pm$10$\cdot10^{-26}$)   \\
1.87$\cdot$10$^{14}$    & 1.6 $\mu$m  & 3.28$\pm0.16\cdot10^{-29}$            &... \\
4.16$\cdot$10$^{14}$    & 7216 $\AA$ & 1.44$\pm0.13\cdot10^{-29}$         &1.03$\pm0.07\cdot10^{-29}$  \\
2.06$\cdot$10$^{15}$    & 1457 $\AA$ &0.15$\pm0.03\cdot10^{-29}$          & 0.054$\pm0.018\cdot10^{-29}$  \\
1.18$\cdot$10$^{17}$    &  soft X   & 1.86$\pm0.8\cdot10^{-32}$           & 0.56$\pm0.27\cdot10^{-32}$ \\
3.39$\cdot$10$^{17}$    & medium X  & 1.04$\pm0.42\cdot10^{-32}$  & $<0.3\cdot10^{-32}$ \\
9.67$\cdot$10$^{17}$    &  hard X  & 0.14$\pm0.14\cdot10^{-32}$       & 0.56$\pm0.28\cdot10^{-32}$ \\
\noalign{\smallskip}
\hline
\end{tabular}\\

Notes to table

{\footnotesize Values in parentheses are uncertain because a
  1.4$^{\prime\prime}$ beamsize is inadequate to isolate S11.3 from
  adjacent knots.}
\end{flushleft}
\end{table}

Using our new Chandra observation we find detections of two knots
(fig.~\ref{fig:evt2}).  The first knot, S3.7, lies at a projected distance of 12.8
kpc from the nucleus and is resolved with the VLA.  The deconvolved
size (FWHM) is 0.46$^{\prime\prime}\times0.18^{\prime\prime}$
(1.6$\times$0.6~kpc) in PA=115$^{\circ}$.  
Following a gap
with no detectable radio emission, the jet again becomes visible in
the radio about 7$^{\prime\prime}$ from the nucleus and brightens as
it approaches the region of maximum apparent curvature.
It is at this point we detect X-rays from the radio knot, S11.3, which
has the highest radio surface brightness (after the nuclear emission).
Its deconvolved radio size is 0.4$^{\prime\prime}\times0.3^{\prime\prime}$
(1.4$\times$1.0~kpc) in PA=48$^{\circ}$.

X-ray fluxes for both knots are given in Table~\ref{tab:xflux} and
flux densities, evaluated for a spectral index equal to 1, are listed
in Table~\ref{tab:fluxden}.  
The total number of counts detected for S3.7 and for S11.3 is 12 and 5 respectively, 
where the average background evaluated around each knot is 1.57 cts for S3.7 and 0.08 cts for S11.3 for the same size aperture.   

\begin{figure}[!htp]
\includegraphics[height=6cm,width=7cm,angle=0]{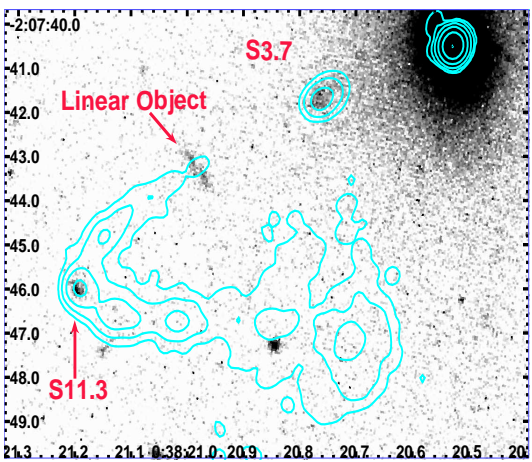}
\caption{HST optical image of 3C~17 (7216\AA). The overlaid cyan
    contours are the same as those in fig.~\ref{fig:evt2}.  Here, the
    radio knots at 3.7$^{\prime\prime}$ and 11.3$^{\prime\prime}$ show
    optical counterparts.  The linear object discussed in the text lies at
    about 7.4$^{\prime\prime}$ from the nucleus.}\label{fig:hst}
\end{figure}

\begin{figure}[!htp] 
\includegraphics[height=8cm,width=8cm,angle=-90]{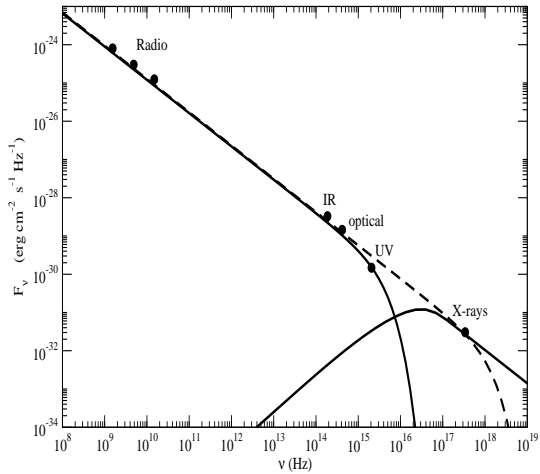}
\caption{The spectrum of S3.7. The synchrotron calculations (done under
the assumptions described in \S\ref{sec:results}) fit the radio to UV with a
cutoff, or extend to the X-ray ignoring the UV datum (dashed line).
The IC/CMB model is the separate component peaking just below
10$^{17}$Hz.}\label{fig:spec37}
\end{figure}

\begin{figure}[!htp] 
\includegraphics[height=8cm,width=8cm,angle=-90]{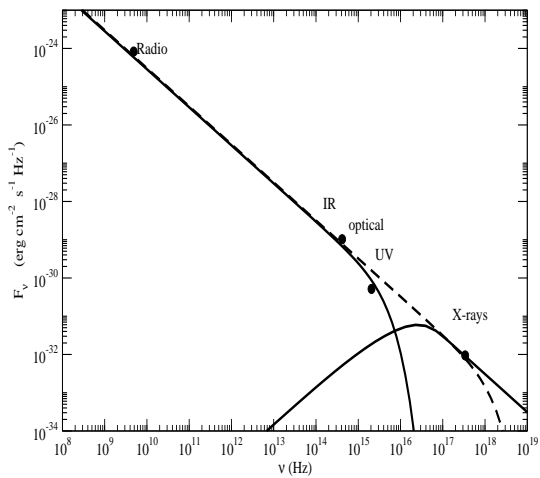}
\caption{The spectrum of S11.3 See the caption of
fig.~\ref{fig:spec37}. }\label{fig:spec113}
\end{figure}

We compared also the radio and the X-ray emission with the optical and
IR images reported by Donzelli et al. (2007).  The knot S3.7 shows a
counterpart both in the optical and in the IR image.  For the other
knot, S11.3, we report only the optical association because the FOV of
the NICMOS camera is too small.  The IR to UV flux densities are
corrected for the reddening using the following values of absorption:
$A_H$=0.01, $A_R$=0.062 and $A_{UV}$=0.044, for the IR, the R band and
the UV respectively (Cardelli et al. 1989)\footnote{see also ``Doug's
Excellent Absorption Law Calculator'',
http://wwwmacho.mcmaster.ca/JAVA/Acurve.html}.  The SED of each knot
is shown in figs.~\ref{fig:spec37}~\&~\ref{fig:spec113}.

We tried several models to fit the emitted spectrum of both knots in
3C~17.  We describe the spectrum from the radio band to the UV in
terms of synchrotron emission but consider both synchrotron and IC/CMB
for the X-rays.  We performed our calculations assuming the following
hypothesis: (1) the distribution of emitting electrons is a power-law
with slope $s$; (2) the volumes of the emitting regions correspond to
the deconvolved radio sizes; (3) the magnetic field is in
equipartition with the energy density of the relativistic electrons; 
and (4) the proton-to-electron ratio is assumed to be zero.

Based on these assumptions, the spectrum is described in terms of 4
parameters, namely: the slope of the electron distribution $s$, the
maximum and the minimum energy of emitting particles
$\gamma_{max},\gamma_{min}$ and the magnetic field, because we fixed
the volume derived from the radio maps.

The spectral index of the electron distribution has been derived
from the observed spectrum fitting the radio to optical data with a
power-law (see Tab.3).  The maximum energy of particles has been
evaluated in order to see the synchrotron exponential cut-off in the
UV, as suggested by the data.  Finally, given the value of the magnetic
field and assuming a minimum observed frequency of 10$^7$Hz, we
derived the $\gamma_{min}$ parameter for both electron distribution
(see Tab. 3).
Following this criteria we found that the $\gamma_{min}$ is the order of 
100 for both knots, it correspond to an electron minimum energy of about 50 MeV, 
that can be predicted by several acceleration processes (e.g. Protheroe 2004).
Finally, the values of the magnetic field, reported in Tab.3, are related to the synchrotron 
interpretation, but are only slightly different for the smaller $\gamma_{min}$.

Solutions shown in figs.~\ref{fig:spec37}~\&~\ref{fig:spec113} have the
beaming factor fixed to 1.  We find a good agreement of our solution
with the observed spectrum for a magnetic field of 180 $\mu$G for S3.7
and 195 $\mu$G for S11.3.  
Parameters for our model are reported in Table~\ref{tab:17sync}. 

For an inverse Compton (IC) model using photons of the microwave
background (CMB) we follow the formalism of Harris \& Krawczynski
(2002).  For S3.7, we find that the required beaming factor is
$\delta$=8 for the fiducial condition $\delta=\Gamma$ ($\Gamma$ is the
bulk Lorentz factor of the jet knot).  The angle to the line of sight
for this solution is $\theta$=7$^{\circ}$ although smaller angles
together with smaller $\Gamma$ are also acceptable.  For S11.3, the
values for the fiducial condition are $\delta=\Gamma$=5.5 and
$\theta$=10$^{\circ}$.  Since the jet is obviously curved, there is no
expectation that the two values of $\theta$ should be equal even if
one might guess that S11.3 would be moving more towards us than S3.7.
These estimates are very rough because of the poor signal to noise in
the X-ray data and because we do not have a measure of the radio
spectral index.

It should be noted that the IC/CMB interpretation rests solely on the
UV flux densities which are approximately 5 sigma for S3.7 and less
than 3 sigma for S11.3 below the single power-law extending to the X-ray data.  
Because of the low statistical significance
of these data and because of the added uncertainty of the extinction
correction, we do not rule out a synchrotron spectrum extending out
to the X-rays as shown by the dotted lines in figures~\ref{fig:spec37}
\&~\ref{fig:spec113}.
However the low UV flux values favor the IC/CMB 
interpretation instead of the single synchrotron component.
A more accurate observation in the UV band (down to $\sim$ 1200$\AA$) is needed 
to distinguish between models, e.g. if the spectrum is curved in the IR-to-UV range 
then the single synchrotron component could be ruled out.

\begin{table}
\caption{Model parameters for the synchrotron calculations.\label{tab:17sync}}
\begin{flushleft}
\begin{tabular}{lll}
\hline
Parameter (cgs units) & S3.7 & S11.3 \\
\hline 
\noalign{\smallskip}
Spectral Index, $\alpha_{ro}$ & 0.87 & 0.99 \\
Electron index $s=2\alpha+1$ & 2.74 & 2.98 \\
$\gamma_{min}$           &  104   &  100  \\
$\gamma_{max}$(IC/CMB)   & 1.8$\cdot10^6$ & 2.2$\cdot10^6$ \\
$\gamma_{max}$(synchrotron)  & 3.1$\cdot10^7$  &  2.8$\cdot10^7$  \\  
Magnetic field (10$^{-6}$ G) & 180 & 195 \\
Volume (10$^{64}$cm$^3$) & 1.81 & 4.21 \\
Luminosity (10$^{43}$ erg~s$^{-1}$) & 1.36 & 0.98\\
\noalign{\smallskip}
\hline
\end{tabular}\\
Notes to table

{\footnotesize a)Two values of $\gamma_{max}$ are listed.  The IC/CMB
  entry is for the case where the synchrotron spectrum cuts off after
  the UV data whereas the synchrotron entry is for a synchrotron
  spectrum extending to the X-ray band.
  
  b) The values of the magnetic field reported are related to the synchrotron interpretation.}
\end{flushleft}
\end{table}

\subsection{A peculiar optical feature}

We noticed a linear optical feature on both HST images: 1.6 $\mu$m and
7216 $\AA$ (fig.~\ref{fig:hst}).  It is a little over an arcsec long
(3.5kpc if at the distance of 3C~17), with a major axis within
11$^{\circ}$ of the perpendicular to the jet at just the point where
the radio emission recommences after the 'gap' following S3.7.  This
feature is $\approx7.3^{\prime\prime}$ (a projected distance of 25kpc)
from the nucleus and has an optical AB magnitude of $\approx$23.  At
7200 $\AA$, there is a 'hole' in the center; one or more pixels are
comparable to the background level.  At 1.6$\mu$m, there is no hole.
If we measure just the outer bits, we find a two-point spectral index,
$\alpha_o$ = 1.7$\pm$0.2.  If we measure the entire object, it is
2.2$\pm$0.2.  There is no evidence of X-ray or radio emission
corresponding to this feature.  Flux densities and upper limits are
given in Table~\ref{tab:linear}.

If the object were an edge on spiral at the same distance of 3C~17, its
absolute magnitude would be -17 which, when coupled with an overall
size of 3.5 kpc would mean it could be classified as a dwarf spiral
(see Schombert et al. 1995).

\begin{table}
\caption{Flux densities for the linear object.\label{tab:linear}}
\begin{flushleft}
\begin{tabular}{ll}
\hline
Frequency & Flux density \\
\hline 
\noalign{\smallskip}
5 GHz          & $<$ 0.5 mJy \\
1.6 $\mu$m  &  4.8$\pm$0.3 $\mu$Jy   \\  
7216$\AA$  & 0.73$\pm$0.04 $\mu$Jy  \\  
1457$\AA$  & $<$ 0.15 nJy \\
1$\times$10$^{18}$Hz  & $<$ 0.1 nJy \\
\noalign{\smallskip}
\hline
\end{tabular}\\
\end{flushleft}
\end{table}

We consider 4 types of possibile explanations for this object.

\begin{itemize}

\item{The object is a foreground or background object (e.g. edge on
spiral) and has nothing whatsoever to do with the jet. A rough
estimate of the probability of the jet crossing a random background
source within the 0.2$^{\prime\prime}$ nuclear region (defined by the
lowered brightness center at 7216$\AA$) is 0.2/360=5$\times10^{-4}$.
However, since we don't know how to estimate the probability that the
(upstream) invisible jet just happens to start converting some of its
power into relativistic electrons and B field at this location, and
since we are wary of {\it a posteriori} probabilities, the 'chance
alignment' hypothesis seems unlikely, but remains a possibility.}

\item{The emitting region arises from the interaction of the jet and
some pre-existing entity (e.g. a large HI cloud, only a part of which
gets ionized and produces free-free emission).  This hypothesis can be
tested with an optical spectrum since the most likely method of
creating optical emission from the interaction of a jet and cold gas
would be via ionization leading to recombination lines and an
optically thin continuum.  Since the observed spectrum is inconsistent
with free-free emission from an ionized gas at $>10^4$K, it would have
to be dominated by emission lines.  The two observed bands correspond
to 1.27-1.36 $\mu$ and 5452-6387 angstroms at the redshift of 3C~17.
Neither of these bands would be expected to contain the more likely
emission lines envisaged by this scenario.}

\item{The emitting region comes from an unknown property of the
jet.  To our knowledge, no other jet exhibits such a narrow band feature
perpendicular to its axis}.

\item{The object is indeed an edge on spiral and is a close companion
of 3C~17.  The jet pierces the center of this galaxy and that is why
the jet begins to be visible at this location.  This jet, like all one
sided jets, is coming 'mostly' towards us: perhaps 10$^{\circ}$ -
30$^{\circ}$ to the l.o.s. for this section of the jet.  Since the
edge on spiral's plane is perpendicular to the plane of the sky, the
actual impacting jet will be close to hitting the plane of the galaxy
at an oblique angle, not coming in at the pole, as it appears in the
projected view.  In any event the probability of hitting an object
whose's center subtends 1 kpc$^2$ (as seen from the SMBH of 3C~17) by
chance is $\frac{1}{4\pi R^2}~\leq~1.3\times10^{-4}$ (again, an {\it a
posteriori} probability).}

\end{itemize}

If we refuse to allow 'intent' (e.g. 'intergalactic engineering'), we
are left with an improbable chance alignment, an interaction with some
pre-existing entity, or some new type of jet-related emission.  
An optical spectrum of this object could eliminate some of these possibilities.

\section{How curved jets can provide evidence for IC/CMB X-ray 
emission}\label{sec:curvature} 

Most jets that have been well studied are relatively straight and the
assumption is normally  made that $\theta$, the angle to the
l.o.s., does not change along the jet.  A curving jet provides an
advantage because a changing viewing angle should be reflected in the
run of the ratio, R, of X-ray to radio intensities differently for
synchrotron and IC emissions.  If, and only if, the X-ray emission is
dominated by IC/CMB emission will the emitting regions closer to the line
of sight than the others display anomalously large values of R.

In X-ray synchrotron jets, R is often a sharply decreasing function
of distance down the jet (``class 1'' e.g. 3C273, Jester et al. 2006),
and we normally ascribe this to a decreasing ability to produce
electrons of the required energies (possibly caused by an increasing
magnetic field strength).  In other straight jets normally thought to
be representative of IC/CMB X-ray emission, the ratio may decrease
smoothly (mimicking the synchrotron) or is sensibly constant down the
jet (``class 2'' e.g. 4C19.44, Schwartz et al. 2007).  Normally these
two possibilities are ascribed to a smoothly decelerating flow
(thereby diminishing the IC component) or a relatively uniform value
for $\Gamma$ which would maintain the effective energy density of the
CMB.

For a curved jet with changing $\theta$, there may be anomalous
changes in synchrotron brightness associated with the effect of
$\theta$ on $\delta$ (the beaming factor), but the ratio of X-ray to
radio emission should be preserved and remain unaffected by changing
$\theta$.  In the IC/CMB scenario, the critical point
is that a smaller $\theta$ will lead to a marked change in R (unlike
the synchrotron case), deviating either from the smoothly decreasing
ratio or from a constant ratio.

The knot S11.3 is brighter than adjacent knots in all 3 bands (radio,
optical, and X-rays).  Is this because it is where the curving jet
lies closest to the l.o.s. or is it just an intrinsically stronger
emitting region which might be caused by a longer path length at a
tangential point?  The longer pathlength possibility will again not
disturb the intrinsic X-ray/radio behavior for either emission
mechanism, so if we can show that S11.3 has an anomalously large
ratio, it will be a strong indicator that the X-ray emission process
is IC/CMB.  

\begin{figure}[!htp]
\includegraphics[height=8cm,width=7cm,angle=-90]{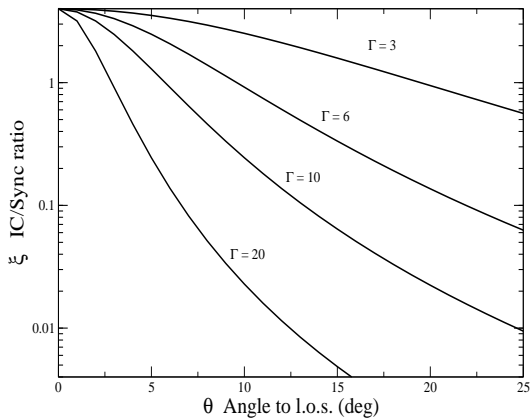}
\caption{The function $\xi$ describing the effects of the extra
  beaming factor on the ratio of IC/CMB to synchrotron emission.
  Curves for four representative values of $\Gamma$ (3, 6, 10, and
  20; top to bottom) are shown.}\label{fig:ratio}
\end{figure}

For IC/CMB emission, the  preference for IC
scattering when the electrons are meeting the photons 'head-on' in
the jet frame, produces more IC emission in the downstream direction.
The angular dependency of this extra beaming term is given in
eq. (A22) of Harris \& Krawczynski (2002) and with a few substitutions
can be written as:

\begin{center}
\begin{equation}
\xi=\left\{1+\frac{\mu\Gamma - \sqrt{\Gamma^2 - 1}}{\Gamma - \mu
    \sqrt{\Gamma^2 - 1}}\right\}^2\end{equation}
\end{center}


where $\Gamma$ is the bulk Lorentz factor of the emitting region and
$\mu=cos \theta$.  This function is shown in fig.~\ref{fig:ratio} for
a few representative values of $\Gamma$.

Although our data for 3C~17 are inadequate to perform a meaningful
test, current parameters are given in Table~\ref{tab:ratio} as an
illustration of the method.  While it is true that R is larger for
S11.3 than for the adjoining knots, R(S3.7) is more than twice
R(S11.3).  To sustain an IC/CMB explanation for the X-rays of both
knots, it becomes necessary to posit a smaller value of $\Gamma$
for the outer parts of the jet compared to that ascribed to S3.7.
That would mean that the $\Gamma^2$ term in eq.(A22) ({\it ibid})
would dominate the change in R between S3.7 and S11.3.  In the
present context the beaming parameters for S3.7 could be (see
\S~\ref{sec:results}) $\Gamma=\delta$=8, $\theta$=7$^{\circ}$ while at
S11.3, $\Gamma$=5.4, $\delta$=5.4, and $\theta$=10$^{\circ}$ and in this case, 
the angular dependence is not the dominant effect.  
Obviously we have too much freedom because of the short \chn~observation: what
is required for this test is a longer observation which would provide
robust X-ray detections of all the knots in Table~\ref{tab:ratio} so as to compare R values
all along the jet.

Other sources with curved jets are 3C120 (for which it is not obvious
where $\theta$ is at a minimum) and 4C65.15, a higher redshift quasar
whose morphology mimics that of 3C~17 and also has an X-ray detection
of a knot at the location of maximum curvature of the jet (Miller \&
Brandt 2008).

\begin{table}
\caption{X-ray to Radio Flux Ratio for the Jet of
3C~17\label{tab:ratio}}
\begin{flushleft}
\begin{tabular}{lrrr}
\hline
Knot & $\nu*S_{5GHz}$     & X-ray Flux & Ratio \\
     & (10$^{-15}$cgs)    & (10$^{-15}$cgs)     &  \\
\hline
\noalign{\smallskip}
S3.7  & 1.45  & 6.4    & 4.4    \\
S10.3 & 1.06  & $<1.6$ & $<$1.5  \\
S11.3 & 3.83  & 7.1    & 1.9  \\
S10.8 & 3.24  & 1.6    & 0.5   \\
S9.6  & 1.80  & $<$1.6 & $<$0.9  \\
S8.0  & 1.25  & $<$1.6 & $<$1.3  \\
S7.2  & 4.63  & $<$1.6 & $<$0.3  \\
\noalign{\smallskip}
\hline
\end{tabular}\\

Notes to table

{\footnotesize The X-ray flux is for the band 0.5 to 7 keV from our
  8ks observation.  Dividing col.3 by col.2 yields the ratios of
  column 4. The X-ray flux of S10.8 comes from a single event and is reported here for illustrative purpose only}

\end{flushleft}
\end{table}

\section{Summary}

We have detected two knots in the 3C~17 jet in both X-rays and
optical/IR bands.  The resulting radio to X-ray spectra do not
provide a definitive answer to the choice between IC/CMB and
synchrotron X-ray emission.   Additionally, we have described a peculiar
optical object, possibly an edge-on spiral galaxy, which appears
to be pierced by the jet.  Although our X-ray data are not sufficient
for the detection of additional jet knots, we have shown how the ratio
of X-ray to radio intensities for the knots of curved jets can be used
as a diagnostic for the X-ray emission process.  IC/CMB emission,
being more tightly beamed than synchrotron emission would be manifest
by a larger ratio for a knot moving closer to the line of sight than
its neighbors.

\acknowledgments 

We thank the anonymous referee for useful comments that led to improvements in the paper. 
We thank R. Morganti for giving us her 5 GHz VLA map of 3C~17.
F. Massaro is grateful to G. Migliori and S. Bianchi for their
suggestions in the \chn~data analysis, A. Siemiginowska for her help
in the use of the \chn~ CIAO data reduction analysis software, and
E. Liuzzo and S. Giacintucci for their suggestions about the radio
data analysis.  This research has made use of NASA's Astrophysics Data
System; SAOImage DS9, developed by the Smithsonian Astrophysical
Observatory; and the NASA/IPAC Extragalactic Database (NED) which is
operated by the Jet Propulsion Laboratory, California Institute of
Technology, under contract with the National Aeronautics and Space
Administration.  
The National Radio Astronomy Observatory is operated by Associated 
Universities, Inc., under contract with the National Science Foundation.
The work at SAO is supported by NASA-GRANT GO8-9114A.

{\it Facilities:} \facility{VLA}, \facility{HST}, \facility{CXO (ACIS)}



\clearpage

\end{document}